# Stop Making Sense of Bell's Theorem and Nonlocality?
# A reply to Stephen Boughn

Federico Laudisa

Department of Human Sciences, University of Milan-Bicocca

Piazza dell'Ateneo Nuovo 1, 20126 Milan, Italy

**Abstract**

In a recent paper on this journal Stephen Boughn argued that quantum mechanics does not require nonlocality of any kind and that the common interpretation of Bell theorem as a nonlocality result is based on a misunderstanding. In this note I argue that the Boughn arguments, that summarize views widespread in certain areas of the foundations of quantum mechanics, are based on an incorrect reading of the presuppositions of the EPR argument and the Bell theorem and, as a consequence, are totally unfounded.

## 1. Introduction

Since the original publication in 1935, the EPR paper does not stop to puzzle physicists and philosophers alike. Even worse for the Bell nonlocality theorem: should the natural world be nonlocal at a certain scale – as many believe – this is increasingly thought to be a possible sign that our beloved theories of space and time might be far from fundamental. Given the effort that might be necessary to follow consistently the latter path, some resistance to such a straightforward reading of the Bell result is reasonable: nevertheless, even those who endorse such an option are not exempt from the obligation of a correct formulation of the state of the art, something which Stephen Boughn fails to provide in his recent paper (Boughn 2017). The present note is intended to show why. Most of the points I will make refer to problems that are far from new, and some of them date back to the very year in which the nonlocality result was published: however, the appearance of claims that appear to forget some basic, well-established points, useful to locate the EPR-nonlocality issue in the logical space that it deserves, makes it essential to revive them.

My note will touch, in the order, (i) the incorrect formulation of the meaning of the Bell theorem, its logical premises and their relation to the EPR argument; (ii) the related role of the so-called counterfactual definiteness condition in the EPR argument that is alleged to carry over in the derivation of the Bell theorem; (iii) the claim that a locality condition which in fact equivalent to what is known as *outcome independence* (OI) is not violated in QM and finally (iv) the plausibility of an ambiguous claim by W.H. Furry concerning the meaning of the original EPR argument, a claim that Boughn defends in support of the idea that entanglement is not peculiar to quantum mechanics.

## 2. 'Classicality' and the Bell theorem

The key point in Boughn's arguments concerns the formulation itself of the Bell theorem that he provides[1]. This is his wording:

Bell's theorem provides a general test, in the form of an inequality, that the results of an EPR type gedanken experiment must satisfy if it is describable *by any classical theory, even one with hidden variables*, so long as such a model is locally causal. Bell then proceeded to demonstrate how the predictions of quantum mechanics violate this inequality. His conclusion was that any hidden variable theory designed to reproduce the predictions of quantum mechanics must necessarily be nonlocal and allow superluminal interactions. (Boughn 2017, p. 2)
[…] Bell's demonstration that any *classical*, *hidden variable* treatment of the system demands acausal, superluminal signals provides sufficient grounds to summarily dismiss such classically based models. (Boughn 2017, p. 3, italics added)

It is quite clear then that, according to the Boughn presentation, the Bell theorem has among its assumption a condition of *classicality*: a (far from clearly defined) condition, that it is reasonable to interpret intuitively as the requirement that any physical system under scrutiny is supposed to be endowed at any given time with a bunch of robust properties, each of which unambiguously characterized and holding independently from any measurement interaction with the system (*pre-existing properties* is a suitable term that has been sometimes used to denote these alleged properties). This appears to be the sense that the term *classicality* is supposed to convey, the sense of a representation of the world in which the physical theory does nothing but 'disclose' a state of affairs that simply is what it is, in total independence from the questions that we scientists decide to raise about it. That this is the intended interpretation is clear from another passage where, with reference to the mathematically precise formulation of the Bell locality, Boughn states:

To be sure, in his 1964 paper, Bell did not conclude that quantum mechanics is nonlocal, only that a classical hidden variable model designed to reproduce the statistical predictions of quantum mechanics must necessarily be nonlocal. However, in a subsequent paper, he formalized "a notion of local causality"

---

[1] The story of this misunderstanding is long: see for instance the recollection in Laudisa 2008, pp. 1113-1116.

that was directly applicable to quantum mechanics and concluded that quantum mechanics itself is nonlocal. This analysis used a mathematical expression of locality similar to the above direct product of probabilities. *The other concept necessary for the derivation of Bell's original inequality was a notion of "classical realism", which in the 1964 paper takes the form of classical hidden variables and the counterfactual definiteness they imply*. (Boughn 2017, p. 11, italics added)

In this vein, a passage from the Bell 1964 paper is invoked, a passage that *verbatim* seems to make the above 'classicality-inspired' readings admissible. It is the final statement of the paper, where Bell summarizes in a sketchy way the result proved in the preceding pages: "In a theory in which *parameters are added to quantum mechanics to determine the results of individual measurements*, without changing the statistical predictions, there must be a mechanism whereby the setting of one measurement device can influence the reading of another instrument, however remote." (Bell 1964, in Bell 2004, p. 20).

The ambiguity in the Bell passage notwithstanding, however, the logic underlying the Boughn claims is not correct. In the opening lines of his 1964 paper, the Bell formulation does not suffer from ambiguities:

The paradox of Einstein, Podolsky and Rosen was advanced as an argument that quantum mechanics could not be a complete theory but should be supplemented by additional variables. These additional variables were to restore causality and locality. In this note that idea will be formulated mathematically and shown to be incompatible with the statistical predictions of quantum mechanics. *It is the requirement of locality, or more precisely that the result of a measurement on one system be unaffected by operations on a distant system with which it has interacted in the past, that creates the essential difficulty*. (Bell 1964, in Bell 2004, p. 14).

To make it short[2], the EPR argument can be formulated as an inference from three conditions to the incompleteness of quantum mechanics: the first is consistency with the statistical predictions of quantum mechanics, the second is the infamous "element-of-physical-reality" condition, whose original formulation in the EPR paper Einstein was dissatisfied with (Howard 1985), and the third is of course locality. It must be stressed that assuming the "element-of-physical-reality" condition is *not* equivalent to assuming the *existence of elements-of-physical-reality* (the 'classicality' Boughn seems to refer to) as an autonomous condition: on the contrary, assuming this condition simply amount to require a *criterion* in order for a property of a physical system to be an objective (i.e. measurement-independent) property[3]. The effective existence of such properties is rather *a consequence* of the locality assumption. Since we start with an entangled state of a composite system, in which spin properties of each EPR subsystem are not elements of physical reality and we end dealing with post-measurement states in which such properties *are* indeed elements of physical reality, the only option open to a *local* description of the whole process is that those properties *were already there*, and this is something we *derive* from our assumption that all physical processes involved in the EPR-preparation-and-measurement procedure must be local. But if such existence of elements-of-physical-reality (the Boughn 'classicality') is *not* assumed at the outset, we cannot dismiss anymore the Bell theorem as a nonlocality result simply by charging it with the accusation of smuggling some 'classical realism' – whatever it is – into the description since the beginning. If this is true, then we must acknowledge that what the Bell theorem is about is nonlocality.

---

[2] The details can be found in Laudisa 2008, pp. 1118-1122.
[3] Redhead 1987, Ghirardi, Grassi 1994, Norsen 2007, Laudisa 2008, Maudlin 2014.

We need no great effort to support this conclusion. We just need to read properly the Bell text that follows immediately the opening lines of his 1964 paper we referred to above, where Bell effectively summarizes the EPR argument:

Consider a pair of spin one-half particles created somehow in the singlet spin state and moving freely in opposite directions. Measurements can be made, say by Stern-Gerlach magnets, on selected components of the spins $\sigma_1$ and $\sigma_2$. If measurement of the component $\sigma_1 \cdot a$, where a is some unit vector, yields the value +1 then, according to quantum mechanics, measurement of $\sigma_2 \cdot a$ must yield the value –1 and vice versa. Now we make the hypothesis, and it seems one at least worth considering, that if the two measurements are made at places remote from one another the orientation of one magnet does not influence the result obtained with the other. Since we can predict in advance the result of measuring any chosen component of $\sigma_2$, by previously measuring the same component of $\sigma_1$, *it follows* that the result of any such measurement must actually be predetermined. Since the initial quantum mechanical wave function does not determine the result of an individual measurement, this predetermination implies the possibility of a more complete specification of the state. (Bell 1964, in Bell 2004, pp. 14–15], italics added)

As should be clear from a fair reading of the Bell original article, the Bell theorem starts exactly from the alternative established by the EPR-Bohm argument—namely, locality and completeness cannot stand together—and goes for the proof that, *whatever form the completability of quantum mechanics might assume*, the resulting theory cannot preserve the statistical predictions of quantum mechanics and be local at the same time: this means that neither a pre-existing-property assumption (or 'Objectivity' or 'Classicality' or whatever synonymous one likes to choose) nor a determinism assumption are assumed in the derivation of the original Bell inequality.

In addition to the logical status of the premises of the Bell theorem, there is a deep motivation for restricting the attention to locality. In the area of investigations opened nearly half a century ago by John S. Bell, the question naturally arose of what would have been the implications of *extending* quantum mechanics, in view of the emergence of phenomena that were not easy to accommodate within a familiar view of the physical world, non-locality being the most urgent case. Due to the unavoidable existence of entangled states – something that makes quantum mechanics a non-local theory in a fundamental sense due to the linearity of the theory, of which entanglement is a consequence (more on this later) – it has seemed plausible to put things in the following way: let us ask whether quantum mechanics might be seen as a 'fragment' of a more general theory which – at a 'higher' level – may recover that locality that turns out not to hold at the strictly quantum level. One of the *strong points* of the original Bell strategy that led to the Bell-named theorem was exactly that this hypothetical extension was confined to the locality/non-locality issue and needed not say *anything* on further details concerning 'realistic' or 'non-realistic' properties, states or whatever: in addition to being useful for the economy of the theorem, this point was absolutely plausible since it makes sense to require from the extension the only condition that we are interested to add to the new hypothetical super-theory, namely locality.

## 3. 'Classicality', counterfactual definiteness and locality

In addition to the alleged role of 'classical realism' we referred to above, a related claim has been around for decades now. According to this claim, the Bell result ensuing from the EPR argument need not be interpreted in terms of nonlocality since it depends on a counterfactual type of reasoning which would be legitimate only in a 'classical' context; for its supporters, this is another motivation for taking the Bell theorem as a 'no-classicality' rather than a nonlocality result.

Boughn in his paper supports this claim:

Bell's theorem clearly makes use of counterfactual definiteness; his inequality involves the correlations of the spins of the two particles in each of two different directions that correspond to non-commuting spin components. This use of counterfactuals is entirely appropriate because it is used to investigate a test for classical hidden variable theories.

What is the logic in this case? Take an ordinary EPR situation with spin 1-2 particles. According to the condition of reality, the choice of the experimenter on one side to measure spin along the *z*-axis allows her to predict with certainty the outcome of the same measurement on the other side and, as a consequence, to conclude that the outcome of this measurement corresponds to an element of physical reality there. The choice *could have been different*, though: the *x*-axis could have been selected as the axis along which the experimenter could perform the spin measurement. The conclusion would have been in this case that the element of physical reality on the other side was the outcome of the measurement along the same axis *x*. Since in fact the experimenter can only perform one measurement at a time, she cannot conclude that *both* the measurement outcomes (along the *z*-axis *and* along the *x*-axis) correspond to elements of physical reality on the other side, but rather only that one *or* the other does. To assume instead that both elements of physical reality are fixed even if quantum mechanics prevents that – so the argument concludes – is essentially equivalent to assume again a form of classicality (the first formulation of this argument is probably due to Peres 1978).

The flaw in this argument is similar to the one discussed in the previous section, namely the supposition that the existence of the elements of physical reality is *added* on top with no other justification than to show that this 'classical-like' supposition cannot hold in a quantum world. If, on the other hand, we take seriously (i) the fact that it is the locality assumption that is prior, and (ii) that it is exactly this assumption that legitimates to talk of elements of physical reality corresponding to spin measurements along different axes, we realize that the above appeal to counterfactual reasoning does no harm. *If* the theory by which we try to account for EPR correlation is assumed to be local, *then* the choice of a certain axis for spin measurement *here* is totally unaffected by what is the chosen axis for spin measurement *there*, and this holds perfectly the same also in a counterfactual sense. Suppose in the actual world the experimenter has chosen the axis *z*: if her measuring operations cannot affect by definition what axis is chosen – and what outcome has been obtained – on the other side, this holds naturally for *whatever choice* and this makes it invalid to claim that in an ordinary EPR argument we have adopted a conterfactual definiteness assumption *in addition* to locality.

## 4. Locality, outcome independence and quantum mechanics

In a section of his paper Boughn focuses on the so-called *principle of separation* (*Trennungsprinzip*), an assumption that Einstein started to use instead of the infamous condition for the elements of physical reality after 1935, unhappy as he was of the formulation of the incompleteness argument for quantum mechanics in the EPR paper (cp. again Howard 1985). The principle (that we will denote by **SEP**<sub>Einst</sub>) reads as follows:

After the collision, the real state of (AB) consists precisely of the real state A and the real state of B, which two states have nothing to do with one another. The real state of B thus cannot depend upon the kind of measurement I carry out on A [**SEP$_{Einst}$**]. But then for the same state of B there are two (in general arbitrarily many) equally justified $\Psi_B$, which contradicts the hypothesis of a one-to-one or complete description of the real states. (Einstein 1935, cit. in Howard 2007)

In fact, the above passage turns out to be an extra-short version of the post-EPR incompleteness argument defended by Einstein, in which completeness is conceived as a one-to-one correspondence between the wave function and the 'real state' of the system (as a matter of fact, this notion of 'real state' is no less mysterious than the EPR reality condition from which Einstein meant to depart). In this version, quantum mechanics is shown to be incomplete because the possibility of different choices of a measurement on one side licenses a plurality of wave functions associated to the 'real state' of tyhe system on the other side, a plurality that violates the above one-to-one correspondence (Howard 2007).

Independently of the fate of the several Einsteinian incompleteness arguments, Boughn proposes an 'experimentalist' reading of the separation principle (let us call it **SEP$_{Exp}$**), according to which *if systems A and B are spatially separated, then a measurement of system A can, in no way, have any effect on any possible measurement of system B* (Boughn 2017, p. 6), and claims that, while the principle **SEP$_{Einst}$** requires "that a measurement of system A can have no effect on the *state* of system B", the principle **SEP$_{Exp}$** requires "that a measurement of system A can have no effect on the *result* of any measurement on system B", a condition that – in Boughn's view – is not violated in standard quantum mechanics (Boughn 2017, p. 6, italics in the original). Some remarks are in order. First, the translation of **SEP$_{Einst}$** into **SEP$_{Exp}$** can hardly be conceived as a gain in conceptual clarity: while the Einsteinian notion of (in)dependence between *states* can be provided with a definite standard meaning, no matter what Einstein might have meant with the expression '*real* state', the same cannot be said of the notion of (in)dependence between *measurements*, if not within a detailed model of the measurement process that Boughn does not mention. Second, this lack of clarity about the (in)dependence between measurements prevents from really grasping what the requirement "a measurement of system A can have no effect on the *result* of any measurement on system B" might amount to. The shift from the term *measurement* to the term *result* in the above requirement may suggest a comparison of the Boughn argument with a distinction that was introduced long time ago, apparently in order to achieve what was known as a 'peaceful coexistence' between quantum mechanics and relativity theory about nonlocality:

the distinction between *parameter independence* and *outcome independence* in the context of stochastic hidden variable models of quantum mechanics (Shimony 1983).

Let us employ the symbol $\lambda$ to denote all parameters useful to characterize the complete specification of the state of an individual physical system (the presentation follows Ghirardi *et al.* 1993). In a standard EPR-Bohm-like situation, the expression

$$P_\lambda^{LR}(x, y; \mathbf{n}, \mathbf{m}) \quad (1.1)$$

denotes the joint probability of getting the outcome $x$ ($x = \pm 1$) in a measurement of the spin component along $\mathbf{n}$ at the left (L), and $y$ ($y = \pm 1$) in a measurement of the spin component along $\mathbf{m}$ at the right (R) wing of the apparatus. We assume that the experimenter at L can make a free-will choice of the direction $\mathbf{n}$ and similarly for the experimenter at R and the direction $\mathbf{m}$. Both experimenters can also choose not to perform the measurement. Bell's locality assumption can be expressed as

$$P_\lambda^{LR}(x, y; \mathbf{n}, \mathbf{m}) = P_\lambda^{L}(x; \mathbf{n}, *) \, P_\lambda^{R}(y; *, \mathbf{m}) \quad (1.2)$$

where the symbol $*$ in the probabilities at the r.h.s denotes that the corresponding measurement is not performed. Jarrett has shown that condition (1.2) is equivalent to the conjunction of two logically independent conditions, namely (Jarrett 1984)

$$P_\lambda^{L}(x; \mathbf{n}, \mathbf{m}) = P_\lambda^{L}(x; \mathbf{n}, *) \quad (1.3a)$$
$$P_\lambda^{R}(y; \mathbf{n}, \mathbf{m}) = P_\lambda^{R}(y; *, \mathbf{m})$$

and

$$P_\lambda^{LR}(x, y; \mathbf{n}, \mathbf{m}) = P_\lambda^{L}(x; \mathbf{n}, \mathbf{m}) \, P_\lambda^{R}(y; \mathbf{n}, \mathbf{m}) \quad (1.3b)$$

Conditions (1.3a) – referred to as *parameter independence* (PI) – jointly express the requirement that the probability of getting a result at L (R) is independent from the setting chosen at R (L), while condition (1.3b) – referred to as *outcome independence* (OI) – expresses the requirement that the probability of an outcome at one wing does not depend on the outcome which is obtained at the other wing. Now we assume – a reasonable move, I think – that we make sense of the Boughn requirement "that a measurement of system A can have no effect on the *result* of any measurement on system B", allegedly presupposed by **SEP**$_{\text{Exp}}$, in terms of OI: under this assumption, however, it can be shown that standard quantum mechanics *does* violate it. For in the standard EPR case when $\lambda$ is the singlet state $\Psi$, if we choose $\mathbf{n} = \mathbf{m}$ we get

$$P_\Psi^{LR}(1, -1; \mathbf{n}, \mathbf{n}) = P_\Psi^{LR}(-1, 1; \mathbf{n}, \mathbf{n}) = 1/2$$
$$P_\Psi^{LR}(1, 1; \mathbf{n}, \mathbf{n}) = P_\Psi^{LR}(-1, -1; \mathbf{n}, \mathbf{n}) = 0$$

but for any *x, y*

$$P_\Psi^L(x; \mathbf{n}, \mathbf{n}) \, P_\Psi^R(y; \mathbf{n}, \mathbf{n}) = 1/4,$$

a result that shows the quantum mechanical violation of outcome independence for certain choices of parameters. This allows us to draw two conclusions: first, if we suppose that the distinction between PI and OI captures more precisely the vague principle behind **SEP**$_{\text{Exp}}$, it is not true that standard quantum mechanics does not contemplate violations of such principle. As a consequence, and this is the second point, standard quantum mechanics already embodies a certain rate of nonlocality *anyway*, even if for the sake of discussion we put into bracket the more general Bell nonlocality theorem (according to which *any theory whatsoever* incorporating the statistical predictions of quantum mechanics is bound to be nonlocal).

## 5. Does entanglement holds in classical physics? On an argument by W.H. Furry

In a further section on the relation between entanglement and nonlocality, Boughn argues that entanglement does not point to a peculiar class of interactions, when he writes that "the phenomenon of entanglement is not restricted to quantum mechanics. Two classical systems that interact with each other before moving off in different directions are also entangled. To the extent that the interaction can be completely characterized, one can predict the correlations of all possible measurements made on the two systems whether space-like separated or not." (Boughn 2017, p. 12). In support of this surprising claim, he employs a highly ambiguous argument of W.H. Furry, contained in a written report of a public debate on the foundations of quantum mechanics which took place at the Xavier University in 1962 and conceived as an illustration of the EPR argument. Let us see the entire Furry 'reworking' of the so-called paradox in terms of a 'classical', macroscopic ordinary situation:

First, you get two envelopes. Then some person, who becomes incommunicado or commits suicide immediately afterwards, takes one or the other of two playing cards, the red or the black, (we don't know which) and tears it in two, and puts half in each envelope. One of the envelopes is sent to Chicago and at any time we can tell what the color of the half card in that envelope in Chicago is just by opening the envelope we have here. We can tell it instantaneously. It doesn't matter if they are opening the envelope in Chicago simultaneously with the one we have here, or before, or after. They will always correlate. This correlation was established in a way that didn't involve any violation of relativity, because they were both together at the time they were put into the envelopes. (C-F-QM 1962, pp. 88-89)

Boris Podolsky, who was also at the debate, rightly replies that "our opening one envelope to determine what the card is in Chicago does not in any way affect the possibilities in Chicago. While in this quantum mechanical experiment, it does, depending on whether we choose to open one envelope or the other" (C-F-QM 1962, p. 89). But Furry reiterates his point:

It's enough to use, say, two envelopes. We enclose them in a slightly infernal box so that the removing of one of these envelopes from the box will promptly result in the complete obliteration of the other one. Now we have two of these boxes, each with two envelopes. The person tears apart a card out of a deck and puts half in each of these two envelopes. For one of them he chooses a card which is either a black suit or a red suit. For the other one he chooses either a low card or a high card. He puts the black or red in the left -hand envelope, the low or high in the right. Then one box is sent to Chicago and the other is kept here. Now you see, there can never be any contradiction if we pull out the black or red and look at it. The other one is destroyed as soon as we pull it out by the infernal arrangement of the box. If we pull out black or red, we now know that if the corresponding envelope is pulled out in Chicago, we know what the answer will be. If the other envelope is pulled out in Chicago, we don't know anything. In any case, however, the sending of the box is perfectly well understood. *There is no contradiction with relativity, and the attaining of information from one place or the other is just what it sounds like. The difference, of course, between the classical and the quantum picture is that the quantum mechanical state does not correspond to this because this nice classical picture of the box with two envelopes is the hidden parameter description and the hidden parameter description is denied in quantum mechanics. But this is the only difference between the two things and there is no difference at all about the questions of information and of distance and time.* (C-F-QM 1962, p. 90, italics added)

The italicized part is the part that Boughn cites with approval, noting that "what distinguishes quantum entanglement from its classical counterpart is simply the superposition of states and the quantum interference it implies." (Boughn 2017, p. 12). But such part is nothing but an early formulation of the very same misunderstanding about the role of 'classicality' we have been discussing in the previous pages: the circumstance that in the infernal box example the card color is a definite property well before the act of pulling in Chicago whereas in the EPR scenario spin properties are *not* definite is a difference that makes the two situations totally irreducible to one another. To defend the opposite is essentially equivalent to arguing that in an EPR scenario we assume at the outset that the spin components are pre-existing properties, a claim that we have previously shown to be incompatible with a correct reading of the EPR-Bell analysis.

# 6. Conclusions

A die-hard tendency in and out of the area of the foundations of quantum mechanics periodically re-emerges, attempting to claim that after all the Bell theorem did not provide a demonstration that the microphysical world, whatever it is, is nonlocal, but simply an elegant rephrasing of a truism, namely that quantum mechanics is not a classical theory! In this note, I argued against the most recent reformulation of this stance, exemplified by the Boughn 2017 paper, by showing that its flaws concern both the logical structure of the nonlocality theorem and its conceptual content.